\documentclass[twocolumn,aps,prb,showpacs,floats,reprint,10pt]{revtex4-1}
\usepackage{amssymb}

\usepackage{amsmath}
\usepackage[T1]{fontenc}
\usepackage{fourier}
\usepackage{graphicx}
\usepackage[colorlinks=true,
allcolors=blue]{hyperref}

\begin{document}

\title{Josephson-like spin current in junctions composed of antiferromagnets
and ferromagnets}
\author{A.~Moor, A.~F.~Volkov, and K.~B.~Efetov}

\affiliation{Institut f\"{u}r Theoretische Physik III,\\
Ruhr-Universit\"{a}t Bochum, 44780 Bochum, Germany\\
}

\begin{abstract}
We study Josephson-like junctions formed by materials with
antiferromagnetic~(AF) order parameters. As an antiferromagnet, we consider
a two-band material in which a spin density wave~(SDW) arises. This could be
Fe\nobreakdash-based pnictides in the temperature interval ${T_{\text{c}}\leq T\leq T_{%
\text{N}}}$, where~$T_{\text{c}}$ and~$T_{\text{N}}$ are the critical
temperatures for the superconducting and antiferromagnetic transitions,
respectively. The spin current~$j_{\text{Sp}}$ in AF/F/AF junctions with a
ballistic ferromagnetic layer and in tunnel AF/I/AF junctions is calculated.
It depends on the angle between the magnetization vectors in the AF leads in
the same way as the Josephson current depends on the phase difference of the
superconducting order parameters in S/I/S tunnel junctions. It turns out
that in AF/F/AF junctions, two components of the SDW order parameter are
induced in the F\nobreakdash-layer. One of them oscillates in space with a short period
${\xi_{\text{F,b}} \sim \hbar v/\mathcal{H}}$ while the other decays monotonously from
the interfaces over a long distance of the order ${\xi_{\text{N,b}}=\hbar v/2\pi T}$~(where~
$v$,~$\mathcal{H}$ and~$T$ are the Fermi velocity, the exchange energy and the
temperature, respectively; the subindex $\text{b}$ denotes the ballistic case). This
is a clear analogy with the case of Josephson S/F/S junctions with a
nonhomogeneous magnetization where short- and long\nobreakdash-range condensate
components are induced in the F\nobreakdash-layer. However, in contrast to
the charge Josephson current in S/F/S junctions, the spin current in AF/F/AF
junctions is not constant in space, but oscillates in the ballistic F\nobreakdash-layer.
We also calculate the dependence of~$j_{\text{Sp}}$ on the deviation from
the ideal nesting in the AF/I/AF junctions. The spin current is maximal in the
insulating phase of the AF and decreases in the metallic phase. It turns to
zero at the Neel point when the amplitude of the SDW is zero and changes sign
for certain values of the detuning parameter.
\end{abstract}

\date{\today}

\maketitle

\section{Introduction}

Similarity between the magnetic and superconducting ordering has been noted
long time ago (see Ref.~\onlinecite{LarkinColeman} and the review articles~\onlinecite{Fomin,Sonin}).
In the simplest case of the BCS superconductivity
the order parameter is described by a complex quantity ${\Delta =|\Delta| \exp(i\chi)}$
with the amplitude~$|\Delta|$ that determines the Cooper
pair density and the phase~$\chi $ related to a voltage~$V$ via the gauge
Josephson equation
\begin{equation}
\hbar \partial_t \chi = q V \,,  \label{I1}
\end{equation}%
with ${q=-2|e|}$, $e$ being the elementary charge. This relation shows that,
in equilibrium, the gauge-invariant quantity~$\mu$ equals zero:
${\mu = \hbar \partial_t \chi - 2 e V}$.

The electrical current~$\mathbf{j}_{q}$ of the condensate is expressed in
terms of another gauge-invariant quantity---the condensate velocity
${\mathbf{v}_{\text{s}} = (\hbar \mathbf{\nabla} \chi - q \mathbf{A}/c)/2m}$
as~${\mathbf{j}_{q} = q n_{\text{s}} \mathbf{v}_{\text{s}}}$.

One of the most remarkable phenomena arising in superconducting systems is
the Josephson effect. Josephson\cite{Josephson} showed that a condensate
current
\begin{equation}
j_{\text{J}} = j_{\text{c}} \sin(\varphi)  \label{I2}
\end{equation}%
can flow in an S/I/S junction being a periodic function of the
gauge-invariant phase difference between superconductors ${\varphi = - (2 m
/ \hbar) \int_{L}^{R} \mathbf{v}_{\text{s}} \, \mathrm{d}\mathbf{l}}$, where
the limits of the integration $R$ and $L$ mean respectively the right and
left superconductors S separated by an insulating layer I, and $j_{\text{c}}$
is the Josephson critical current.

In the case of a magnetic ordering---ferromagnetic~(F) or
antiferromagnetic~(AF)---the order parameter is the magnetization~$\mathbf{M%
}$ (in the antiferromagnets the vector~$\mathbf{M}$ is the magnetization in
one of the sublattices). The absolute value of~$\mathbf{M}$ corresponds to
the amplitude~$|\Delta|$ in superconductors, and the angle~$\theta$
between the $z$\nobreakdash-axis and the vector~$\mathbf{M}$ is analogous,
to some extent, to the phase~$\chi$. The temporal variation of $\theta$ is
described by the equation similar\cite{Flavio,Tremblay} to Eq.~(\ref{I1}),
\begin{equation}
\partial_t \theta = -g \mu_{\text{B}} \delta \mathbf{B} \cdot
(\mathbf{M}_{\text{R}} \times \mathbf{M}_{\text{L}}) / M_{\text{R}} M_{\text{L}} \,, \label{I3}
\end{equation}%
where~$g$ is the gyromagnetic ratio, $\mu_{\text{B}}$ is the Bohr magneton,
${\delta \mathbf{B} = \mathbf{B}_{\text{R}} - \mathbf{B}_{\text{L}}}$ and $%
\mathbf{B}_{\text{R},\text{L}}$ is the magnetic induction in the
respectively right or left F- or AF\nobreakdash-layer composing the
Josephson-like structure. In Eq.~(\ref{I3}), the $z$\nobreakdash-axis is
chosen to be directed along the vector~$\delta \mathbf{B}$.\footnote{We do not discuss Eq.~(\ref{I3}) in detail since
we will study the dc Josepshon-like effect only. More detailed discussion of
the nonstationary relation,  Eq.~(\ref{I3}), is given in Refs.~\onlinecite%
{Flavio,Tremblay}.}

The Josephson-like effect in F/I/F and AF/I/AF junctions was studied in
Refs.~\onlinecite{Lee,Flavio,Tremblay}. It was found that in these
junctions, the spin current has the form
\begin{equation}
j_{\text{Sp}} = j_{\text{c,Sp}} \sin(\alpha ) \,,  \label{I4}
\end{equation}%
where ${\alpha = \theta_{\text{R}}-\theta _{\text{L}}}$ and~$j_{\text{c,Sp}}$ is a
constant related to the barrier transmittance~$|T|^{2}$.

In the full analogy to the Josephson charge current~$j_{\text{J}}$, the
spin current is nondissipative. The absence of the dissipation for the spin
current and the possibility to use this dissipationless current in
spintronics has been discussed in many
papers\cite{Hirsch,Brataas,WangW,Slonczewsi,MacDonald}
(see also reviews~\onlinecite{Prinz,Sarma}).
Unfortunately, using the spin currents in
spintronics does not seem to be straightforward because the analogy
between the spin and charge currents is not complete, namely, the spin
current is not a conserved quantity.\cite{LarkinColeman,Fomin,Sonin} We will
demonstrate the validity of this statement considering the proximity effect
in AF/F structures. In particular, we will calculate the spin current
through a Josephson-like AF/F/AF junction.

It is well known that the Josephson effect in S/F/S junctions reveals new
remarkable properties\cite{GolubovRMP,BuzdinRMP,BVErmp,PhysToday} that are
absent in the case of more conventional S/I/S or S/N/S junctions.

First, the condensate functions~$f$ penetrating into a diffusive F%
\nobreakdash-layer decay on a very short range of the order ${\xi _{\text{F,d%
}}=\sqrt{D\hbar /\mathcal{H}}}$, where~$D$ is the classical diffusion
coefficient of the ferromagnet and~$\mathcal{H}$ is the exchange field; the
subindex~$d$ denotes the diffusive case. This expression for~${\xi_{\text{F,d}%
}}$ is valid if the condition ${\tau \mathcal{H}\ll 1}$ is fulfilled, where
$\tau$ is the momentum relaxation time. Usually, the exchange field~$\mathcal{H}$
is much larger than the critical temperature of the
superconducting transition~$T_{\text{c}}$. In addition, the Cooper pairs
wave function oscillates in space. It consists of the singlet component~$f_{3}$,
and the triplet component~$f_{0}$ with zero projection of the total
spin on the direction of the magnetization~$\mathbf{M}$. Such oscillations
lead to a change of the sign of the Josephson critical current~$j_{\text{c}}$,
and the so called $\pi$\nobreakdash-state is realized in the junction.

Another interesting effect predicted in Ref.~\onlinecite{BVE01} (see also the
reviews~\onlinecite{BVErmp,PhysToday}) and observed in several experiments \cite%
{Klapwijk,Sosnin,Birge,Chan,BlamireScience,Westerholt,Aarts,Deutscher} in
S/F structures, is related to a new type of superconducting correlations.
This type of correlations called ``odd triplet superconductivity'' can arise
due to the proximity effect in the F\nobreakdash-layer with a nonhomogeneous
magnetization. In the case of a uniform magnetization in the F\nobreakdash%
-layer, the pair wave function there has both the singlet component ${f_{%
\text{sng}} \sim \langle \psi_{\uparrow}(t) \psi_{\downarrow}(0) -
\psi_{\downarrow}(0) \psi_{\uparrow}(t) \rangle}$ and the triplet component
${f_{0} \sim \langle \psi_{\uparrow}(t) \psi_{\downarrow}(0) +
\psi_{\downarrow}(0) \psi_{\uparrow}(t) \rangle}$ with zero projection of the
total spin of the Cooper pair on the direction of the magnetization. Both
components oscillate in space and decay on a short length of the order
of~$\xi_{\text{F,d}}$. In the ballistic case, these components oscillate with
the period~${\sim \xi_{\text{F,b}}=v\hbar /2\mathcal{H}}$.

The situation changes drastically if the magnetization in F is not uniform
in the nearest vicinity of the S/F interface.\cite{BVE01,BVErmp,PhysToday}
In this case, not only the components~${f_{\text{sng}} \equiv f_{3}}$ and~$f_{0}$
arise in the F\nobreakdash-layer, but also the triplet
component~${f_{1}(t) \sim \langle \psi_{\uparrow}(t) \psi_{\uparrow}(0) +
\psi_{\uparrow }(0) \psi_{\uparrow}(t) \rangle}$ with nonzero projection of the
total spin on~$\mathbf{M}$ is generated. This component is insensitive to
the exchange field and penetrates the F\nobreakdash-layer over a long
(compared to~$\xi _{\text{F}}$) distance. The penetration length does not
depend on the exchange field~${\mathcal{H}}$ and is much larger than~$\xi_{\text{F}}$.
In absence of the spin-orbit interaction and scattering by
magnetic impurities this length may become comparable with the penetration
length of the superconducting condensate into the diffusive normal~(nonmagnetic)
metal ${\xi_{\text{N,d}} = (D \hbar /2\pi T)^{1/2}}$.

The component~$f_{1}$ can be called the long\nobreakdash-range triplet component~(LRTC).
It is symmetric in the momentum space and is therefore also insensitive to
the scattering by ordinary~(nonmagnetic) impurities. In this respect the
LRTC differs drastically from the usual triplet component~$f_{\text{tr}}$
that describes superconductivity in~$\text{Sr}_{2}\text{RuO}_{4}$.\cite{Maeno}
The latter is an antisymmetric function of the momentum~$\mathbf{p}$
and is therefore destroyed by ordinary impurities. The correlator~$f_{1}$
changes sign by permutation of the operators~$\psi_{\uparrow}(t)$
and~$\psi_{\uparrow}(0)$, as it should be according to the Pauli principle,
which follows from the fact that the Fermi operators~$\psi_{\uparrow}(t)$
and~$\psi_{\uparrow}(0)$ anticommute at equal times. This means that
${f_{1}(0) \sim \langle \psi_{\uparrow}(0) \psi_{\uparrow}(0) +
\psi_{\uparrow}(0) \psi_{\uparrow}(0) \rangle = 0}$, i.e., the function~$f_{1}(t)$
is an odd function of time~$t$ or, in other words, it is an odd
function of the frequency~$\omega $ in the Fourier representation. That is
why this component is called ``odd triplet component''.

Recent experiments unambiguously confirmed the existence of the LRTC in the S/F
structures with a nonhomogeneous magnetization.\cite%
{Klapwijk,Sosnin,Birge,Chan,BlamireScience,Westerholt,Aarts,Deutscher}

In this paper we show that the analogy between the superconducting and
magnetic order parameters in Josephson-like junctions is deeper than it was
indicated previously. It turns out that in the SDW/F/SDW junction with a
ballistic F\nobreakdash-layer the ``short''- and long\nobreakdash-range AF order parameters are
induced in the F\nobreakdash-layer in an AF/F structure (we use the quotation
marks for ``short''-range, to indicate that in the considered ballistic case this
component does not decay on a short range, but rather oscillates with a
short period). The component of the AF order parameter with the $\mathbf{M}$~vector
(in each sublattice) being parallel to the magnetization
vector in the F\nobreakdash-layer~$\mathbf{M}_{\text{F}}$, which is supposed
to be oriented along the $z$\nobreakdash-axis, penetrates into the F\nobreakdash-layer
and oscillates with a short period of the order of~$\xi_{\text{F,b}}$.
On the other hand, the SDW with the $\mathbf{M}$~vector perpendicular
to~$\mathbf{M}_{\text{F}}$ penetrates the F\nobreakdash-layer over a long
distance of the order of~$\xi_{\text{N,b}}$ and decays monotonously. This
part of the SDW order parameter may be called the long\nobreakdash-range
component induced in the ferromagnetic layer.

At the same time, in spite of the long\nobreakdash-range penetration of the AF
correlations into the ferromagnet~F, the spin current that arises at the
AF/F interface is not constant in space. It oscillates in the F region with
a short period. The study of the spin current in various heterostructures is
important not only from the point of view of fundamental physics but also
from the point of view of possible applications of these structures in
spintronics.\cite{Sarma}

We do not analyze the case of the Bose--Einstein condensation in the magnon system at a high magnon density which can be realized via the external pumping. In this case, a Josephson-like effect may also arise, but it has different properties~(see the review~\onlinecite{Volovik} and references therein).

\section{Model}

In order to describe the AF order we adopt the model developed for
superconducting Fe\nobreakdash-based pnictides.\cite%
{Chubukov09,Eremin10,Chubukov10,Tesanovic,Schmalian10} In this model the
band structure is assumed to consist of two bands~(electron and hole) with
perfect or almost perfect nesting, which is taken into account by a certain
parameter~$\delta \mu$. Due to this, a spin density wave~(SDW) corresponding to an
antiferromagnetic order parameter arises in these materials. On the basis of
this model, which describes well many properties of this new class of
high\nobreakdash-$T_{\text{c}}$ superconductors (see review articles~\onlinecite%
{Mazin,Norman,Wilson,Johnston,Stewart,Chubukov11Rev} and references
therein), the generalized Eilenberger equations for the quasiclassical
Green's functions~$\check{g}$ have been derived in the previous paper\cite%
{MoorVE11} by the authors (see also a recent paper~\cite{Chubukov11}
where the definition of the Green's functions differs from ours).
The quasiclassical formalism is especially
suitable for tackling nonhomogeneous problems and the matrix functions~$\check{g}$
allow one to find all the necessary observable quantities, such
as the charge and spin currents, the density-of-states, etc. The technique of
quasiclassical Green's functions is presented, e.g., in Refs.~\onlinecite%
{Rammer,BelzigRev,Kopnin}.

\section{SDW/F/SDW ``Josephson'' junction}

\subsection{``Short''- and Long-range Components}

\begin{figure}[t]
\begin{center}
\includegraphics[width=0.3\textwidth]{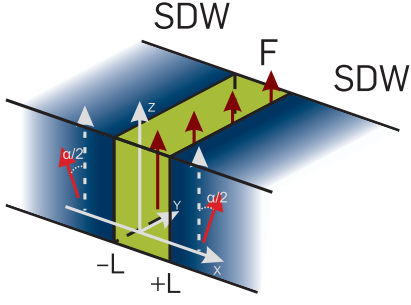}
\end{center}
\caption{{\protect\small {(Color online.) The considered setup. The two
antiferromagnets SDW are separated by a thin ferromagnetic layer of
thickness~$2L$. The orientation of the magnetization vectors of the SDW in
the leads lying in the $y$-$z$-plane is shown in red.}}}
\label{fig:Setup}
\end{figure}

First, we consider a Josephson-like AF/F/AF junction composed of two leads
and a thin ferromagnetic layer between the leads. In the leads a spin density
wave exists, which is a particular example of the AF order and the
layer between these materials is a ferromagnet in the ballistic regime~(see
Fig.~\ref{fig:Setup}).
For simplicity, we assume that both the SDW and the F parts of the structure
consist of similar two-band materials and differ from each other by
different interaction constants and doping levels. The AF/F interface
penetrability is supposed to be small and thus one can neglect the effect of
the F\nobreakdash-layer on the SDW leads. The Green's functions in the
right~(left) leads are given by the formula that can be obtained from
Eq.~(6.4) of Ref.~\onlinecite{MoorVE11} by setting the superconducting order
parameter to zero,~${\Delta =0}$)
\begin{align}
\check{g}_{\text{R(L)}} = \frac{1}{\mathcal{E}_{M}} \big\lbrace \omega_{n} \hat{\tau}_{3}&
+ W_{M0} \big[ \hat{\rho}_{1} \cdot \hat{\tau}_{2} \cdot \hat{\sigma}_{3} \cos(\alpha /2)  \label{M1} \\
& \pm \hat{\rho}_{2} \cdot \hat{\tau}_{1} \cdot \hat{\sigma}_{2} \sin(\alpha /2) \big] \big\rbrace \,.  \notag
\end{align}%
Here, ${\omega_{n} = \pi T (2n+1)}$ is the Matsubara frequency, $W_{M0}$~is
the amplitude of the SDW, ${\mathcal{E}_{M} = \sqrt{\omega _{n}^{2}+W_{M0}^{2}}}$
and ${\hat{\rho}_{i} \cdot \hat{\tau}_{j} \cdot \hat{\sigma}_{k}}$ is the
tensor product of Pauli matrices in the band, Gor'kov--Nambu and spin space,
respectively. The magnetization of the SDW lies in the
$y$\nobreakdash-$z$\nobreakdash-plane at the angle $\pm \alpha /2$ relative to the
$z$\nobreakdash-axis, with the projection on the $y$\nobreakdash-axis
${\pm W_{M0} \sin(\alpha /2)}$ in the left~(right) leads. For simplicity, we
suppose the ideal nesting and set ${\delta \mu =0}$. In the F\nobreakdash-layer,
the quasiclassical Green's functions obey the equation
\begin{equation}
n_{x} v \partial_{x} \check{g} + \left[ \check{\Lambda} \,, \check{g}\right] =0 \,,
\label{M2}
\end{equation}%
where ${n_{x}=p_{x}/p}$ and $v$~is the modulus of the Fermi velocity. The matrix
$\check{\Lambda}$ is defined as ${\check{\Lambda}=\hat{\tau}_{3} \cdot (\omega _{n}\hat{\rho}_{0}
\cdot \hat{\sigma}_{0} + \mathrm{i} \mathcal{H} \hat{\rho}_{0} \cdot \hat{\sigma}_{3})}$
with the exchange field in the ferromagnet~$\mathcal{H}$. We assume that there is no
impurity scattering in the F\nobreakdash-layer~(ballistic case). The impurity
scattering leads to a suppression of the SDW amplitude.\cite{Tugushev,Chubukov11}

Equation~(\ref{M2}) is supplemented by the boundary condition which we write in
the form
\begin{equation}
(\check{g}(n_{x}) - \check{g}(-n_{x})) = \pm \mathrm{sgn}(n_{x})
[\check{g} \,, \check{\mathcal{T}} \cdot \check{g}_{\text{R(L)}} \cdot
\check{\mathcal{T}}] \,.
\label{M3}
\end{equation}%
The transmission coefficient~$\check{\mathcal{T}}$ is a matrix, which is
assumed to be small. This boundary condition generalizes the widely used
boundary condition derived by Zaitsev\cite{Zaitsev} to the case of a
two\nobreakdash-band metal with an SDW. We consider a simplified model for
the AF/F interface and assume that there are no interband transitions. In this
case, the matrix~$\check{\mathcal{T}}$ has the form
${\check{\mathcal{T}}=(\mathcal{T}_{3} \hat{\rho}_{3} + \mathcal{T}_{0}
\hat{\rho}_{0}) \cdot \hat{\tau}_{0} \cdot \hat{\sigma}_{0}}$,
where $\mathcal{T}_{0,3} = (T_{1} \mp T_{2})/2$ and~$T_{1,2}$ are the
matrix tunneling elements for transitions from the band~$1$ resp.~$2$ of the
F\nobreakdash-layer to the same bands~$1$ resp.~$2$ of the SDW leads~(see
the Appendix~\ref{Boundary_Conditions}). If these elements are equal,
i.e., ${T_{1} = T_{2} \equiv T_{\text{tun}}}$, then ${\mathcal{T}_{0}=0}$
and ${\mathcal{T}_{3}=T_{\text{tun}}}$.

The boundary conditions that describe real materials are more complicated.
They must include the transitions between different bands and this process
may be accompanied by spin flips. However, we are not interested in exact
calculations of the amplitude of the SDW penetrating into the ferromagnetic
layer, but rather in a qualitative effect---the appearance of ``short''- and
long\nobreakdash-range components. The existence of these
components does not depend on the exact form of the boundary conditions.

In the lowest order of approximation in the transmission coefficient~$\mathcal{T}$
(no proximity effect), the matrix~$\check{g}$ in the ferromagnet has the
form ${\check{g} \equiv \check{g}_{0\text{F}} = \mathrm{sgn}(\omega )
\hat{\rho}_{0} \cdot \hat{\tau}_{3} \cdot \hat{\sigma}_{0}}$.
Therefore, the boundary condition can be written in the form
\begin{equation}
(\check{g}(n_{x})-\check{g}(-n_{x})) = \pm \mathrm{sgn}(n_{x})
\mathcal{T}_{\text{eff}}^{2} \big[ \hat{\tau}_{3} \,, \check{f}_{\text{R(L)}} \big] \,,  \label{M4}
\end{equation}%
where the matrix~$\check{f}_{\text{R(L)}}$ equals
\begin{equation}
\check{f}_{\text{R(L)}} = \frac{W_{M0}}{\mathcal{E}_{M}} \big\lbrace \hat{\rho}_{1} \cdot
\hat{\tau}_{2} \cdot \hat{\sigma}_{3} \cos(\alpha /2) \pm \hat{\rho}_{2} \cdot
\hat{\tau}_{1} \cdot \hat{\sigma}_{2} \sin(\alpha /2) \big\rbrace  \label{M5}
\end{equation}%
and
\begin{equation}
\mathcal{T}_{\text{eff}}^{2} = -(\mathcal{T}_{0}^{2} - \mathcal{T}_{3}^{2}) = T_{1}T_{2} \label{eqn:eff_transm_coeff}
\end{equation}
is an effective transmission coefficient for the SDW.
The boundary condition~(\ref{M4}) defines the spin current through the
interface. One can see that if one of the tunneling elements~($T_{1}$ or~$T_{2}$)
is zero, then the spin current turns to zero.

We need to find a solution of Eq.~(\ref{M2}) supplemented with the boundary
condition~(\ref{M4}). In the considered case of a low penetrability of the SDW/F interface,
the matrix~$\check{g}$ in the F\nobreakdash-film can be represented in the
form ${\check{g} = \check{g}_{0\text{F}} + \delta \check{g}}$, where~${\delta
\check{g}}$ is a matrix with a small ``amplitude'' which describes the SDW
induced in the F\nobreakdash-film due to the proximity effect. As usual, we
represent~$\check{g}$ as a sum of a symmetric in momentum part and the antisymmetric
one, i.e., ${\delta \check{g} = \check{s} + n_{x} \check{a}}$. We substitute this
expression into Eq.~(\ref{M2}) and split it into two equations for the
symmetric and antisymmetric parts
\begin{align}
v \partial_x \check{s} + \big[ \check{\Lambda} \,, \check{a} \big] &= 0 \,, \label{M2a} \\
n_{x}^{2} v \partial_x \check{a} + \big[ \check{\Lambda} \,, \check{s} \big] &= 0 \,.  \label{M2b}
\end{align}%
Excluding the matrix~$\check{a}$, we obtain one equation for the matrix~$\check{s}$
\begin{equation}
-v^{2} n_{x}^{2} \partial^{2}_x \check{s} + \big[\check{\Lambda} \,, \big[ \check{\Lambda} \,, \check{s} \big] \big] = 0 \,.  \label{M6}
\end{equation}%
For the matrix~$\check{a}$ we obtain a similar equation. The commutator
${\big[ \check{\Lambda} \,, \big[ \check{\Lambda} \,, \check{s} \big] \big]}$ is easily
calculated resulting in
\begin{align}
\big[ \check{\Lambda} \,, \big[ \check{\Lambda} \,, \check{s} \big] \big]
= 2 \big\lbrace (\omega^{2} - {\mathcal{H}}^{2}) \check{s}
&+ 2 \mathrm{i} \mathcal{H} \omega
( \hat{\sigma}_{3} \cdot \check{s} + \check{s} \cdot \hat{\sigma}_{3})  \\
&+ \omega^{2} \check{s} - \mathcal{H}^{2} \hat{\sigma}_{3} \cdot \check{s} \cdot \hat{\sigma}_{3} \big\rbrace \,. \notag
\end{align}
Here, we used the property ${\check{s} \sim \check{\tau}_{1,2}}$.
It is seen from the boundary conditions~(\ref{M4}--\ref{M5}), that this
matrix anticommutes with~${\check{\tau}_{3}}$. It is convenient to represent
the matrix~$\check{s}$ as a sum of two matrices
${\check{s}=\check{s}_{\parallel} + \check{s}_{\perp}}$, where
${\check{s}_{\parallel} = \hat{s}_{0} \cdot \hat{\sigma}_{0} +
\hat{s}_{3} \cdot \hat{\sigma}_{3}}$ and ${\check{s}_{\perp} =
\hat{s}_{2} \cdot \hat{\sigma}_{2}}$. We will see that the matrix
$\check{s}_{\parallel}$ describes a ``short''\nobreakdash-range component
of the SDW penetrating into the F\nobreakdash-layer, while $\check{s}_{\perp}$
determines the long\nobreakdash-range component, which is perpendicular to the
orientation of the magnetization in the ferromagnet. From Eq.~(\ref{M6}) we obtain
the equations for~$\check{s}_{\parallel}$ and~$\check{s}_{\perp}$
\begin{align}
-v^{2} n_{x}^{2} \partial^{2}_x \check{s}_{\parallel} +
4 (\omega + \mathrm{i} \mathcal{H} \hat{\sigma}_{3})^{2} \check{s}_{\parallel} &= 0 \,,  \label{M8a} \\
-v^{2} n_{x}^{2} \partial^{2}_x \check{s}_{\perp} + 4 \omega^{2} \check{s}_{\perp} &= 0 \,.  \label{M8b}
\end{align}

In the first order in the transmission coefficient~$\mathcal{T}_{\text{eff}}^{2}$
the solutions of these equations obeying the boundary conditions are
\begin{align}
\check{s}_{\parallel} &= 2 m s_{\omega} \mathcal{T}_{\text{eff}}^{2}
\hat{\rho}_{1} \cdot \hat{\tau}_{2} \cdot \big[ \hat{\sigma}_{3} \cdot \mathcal{R}(x) +
\mathrm{i} \hat{\sigma}_{0} \cdot \mathcal{I}(x) \big] \cdot \cos (\alpha/2) \,, \label{M9a} \\
\check{s}_{\perp} &= 2 m s_{\omega} \mathcal{T}_{\text{eff}}^{2} \hat{\rho}_{2} \cdot
\hat{\tau}_{1} \cdot \hat{\sigma}_{2} \cdot \mathcal{C}(x) \cdot \sin (\alpha/2) \,, \label{M9b}
\end{align}
where ${s_{\omega} \equiv \mathrm{sgn}(\omega)}$ and ${m = W_{M0}/\mathcal{E}_{M}}$ is
the dimensionless amplitude of the SDW induced in the F\nobreakdash-layer and
\begin{align}
\mathcal{R}(x) &= \Re \left\lbrace \mathcal{J}(x) \right\rbrace \,, & \mathcal{I}(x) &= \Im \left\lbrace \mathcal{J}(x) \right\rbrace \,,\label{M10a} \\
\intertext{and}
\mathcal{J}(x) &= \frac{\cosh(\theta_{+} x/L)}{\sinh(\theta_{+})} \,, & \mathcal{C}(x) &= \frac{\sinh(\theta_{\omega} x/L)}{\cosh(\theta_{\omega})} \,,  \label{M10b}
\end{align}
with ${\theta_{\omega} \equiv \varkappa_{\omega} L}$ and ${\theta_{+} = \varkappa_{+} L}$.
The here introduced wave vectors ${\varkappa_{+} = 2 (|\omega| + \mathrm{i} \mathcal{H} \mathrm{sgn}(\omega)) / v |{n_{x}|}}$
and ${\varkappa_{\omega} = 2 |\omega| / v |n_{x}|}$ determine the characteristic
lengths which describe the penetration of the ``short''- and long\nobreakdash-range
components of the SDW into the F\nobreakdash-region, respectively. The same wave vectors characterize
the penetration of the short\nobreakdash-range component, i.e., the singlet and triplet
component with zero projection of the total spin onto the $z$\nobreakdash-axis, and the
long\nobreakdash-range triplet component of the superconducting wave function into the
F\nobreakdash-region of an S/F bilayer in the ballistic case.\cite{BuzdinRMP,BVErmp,PhysToday}

The component of the SDW~$\check{s}_{\parallel}$, Eq.~(\ref{M8a}),
which can be called the ``short''-range component, oscillates at a given angle
${\chi = \arccos(n_{x})}$ rather fast in space with a period of the order
of~$\xi_{\text{F,b}}$~(it is assumed that ${\mathcal{H} \gg T}$). Being
averaged over the angle, it decays from the interface in a power\nobreakdash-law
fashion. The long\nobreakdash-range component~$\check{s}_{\perp}$, Eq.~(\ref{M8b}),
penetrates into the ferromagnet over a long length~$\sim \xi_{\text{N,b}}$, which
does not depend on the exchange field~$\mathcal{H}$. Thus, there is an
analogy with an S/F/S Josephson junction where the ferromagnet has an
inhomogeneous magnetization.\cite{BVErmp,PhysToday} In the ballistic case, the
``short''\nobreakdash-range component of the condensate wave function oscillates in space
with a period~${\sim \xi_{\text{F,b}}}$,\cite{BuzdinRMP,BVErmp,PhysToday}
whereas the long\nobreakdash-range component penetrates the ferromagnet over
the length~${\sim \xi_{\text{N,b}}}$. We use quotation marks for the
``short''\nobreakdash-range component because, in the considered ballistic case,
this component does not decay exponentially in the ferromagnetic layer, but
oscillates and decays in a power\nobreakdash-law fashion after averaging over the
angles. In the diffusive limit in the S/F/S junctions, this component decays
exponentially on a short distance of the order~$\xi_{\text{F,d}}$.

Note that the ``short''-range component~$\check{s}_{\parallel}$ consists of
the odd and even in~$\omega$  parts (correspondingly, the first and the second terms in Eq.~(\ref{M9a})), whereas the
long-range component~$\check{s}_{\perp}$ is an odd function of
the frequency~$\omega$. The same dependence has an anomalous~(Gor'kov's) Green's function in an S/F/S system with a nonhomogeneous
magnetization~$\mathbf{M}$.\cite{BuzdinRMP,BVErmp} In the last case, the
short-range part consists of the singlet component and triplet component
with zero projection of the total spin onto the~$\mathbf{M}$ vector, and the long-range part consists of the triplet component with a non-zero projection of the total spin onto the~$\mathbf{M}$ vector.

\begin{figure}[t]
\begin{center}
\includegraphics[width=\columnwidth]{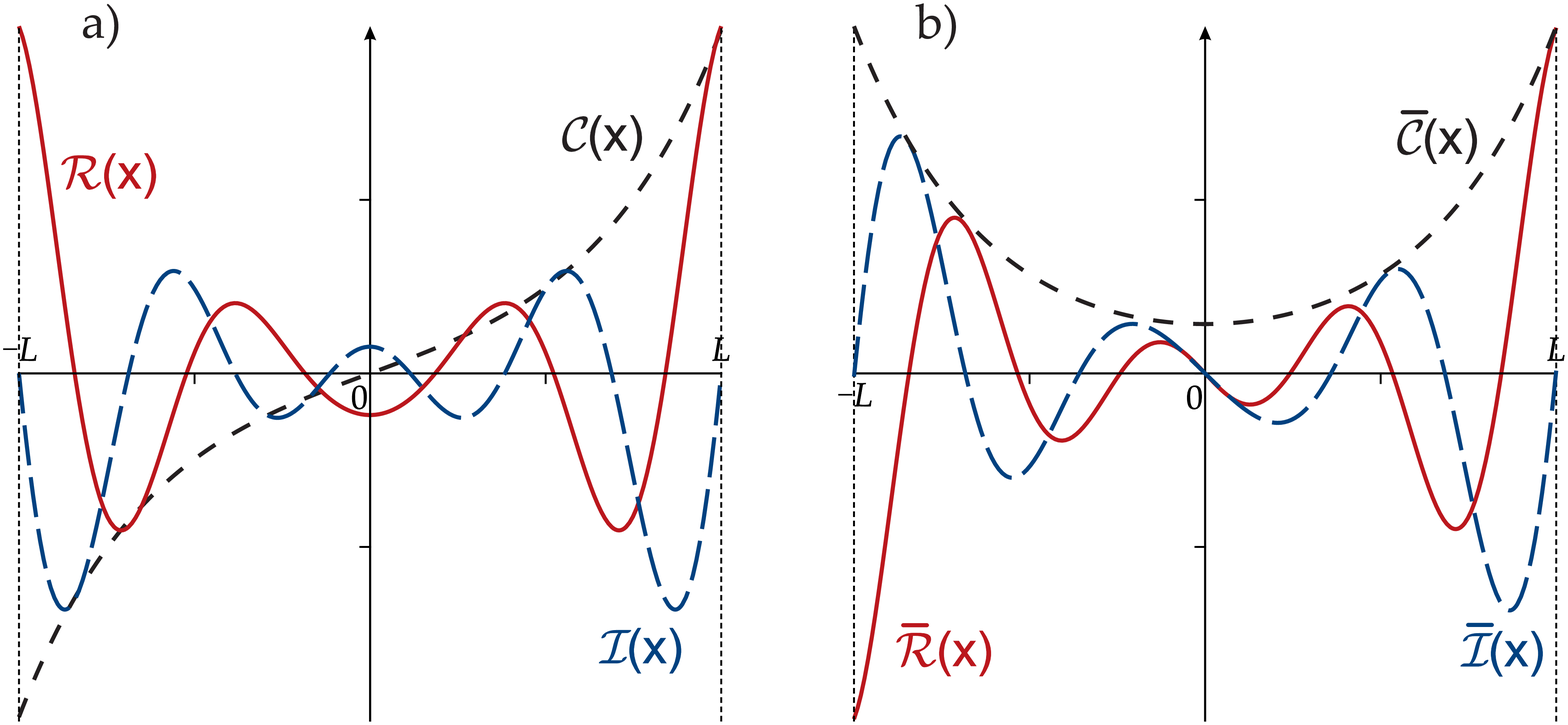}
\end{center}
\caption{{\protect\small {(Color online.) The coordinate dependence of the
``short''-range resp.\ the long-range components of the Green's function in the F-layer.\\
a) the symmetric in momentum part of $\check{g}$---$\mathcal{R}(x)$ and $\mathcal{I}(x)$ resp.~$\mathcal{C}(x)$\\
b) the antisymmetric in momentum part of $\check{g}$---$\bar{\mathcal{R}}(x)$ and $\bar{\mathcal{I}}(x)$ resp.~$\bar{\mathcal{C}}(x)$
}}}
\label{fig:Re_Im_C}
\end{figure}

In Fig.~\ref{fig:Re_Im_C}~a) we plot the coordinate dependence of the ``short''-~(the
functions~$\mathcal{R}(x)$ and~$\mathcal{I}(x)$) and the long\nobreakdash-range~(the
function~$\mathcal{C}(x)$) components. The former stems
from the part of the matrix~$\check{g}$ corresponding to the magnetization
component of the SDW~$\mathbf{M}$ parallel to the magnetization in the
ferromagnet~$\mathbf{M}_{\text{F}}||\mathbf{z}$, i.e., the second term on
the right in~Eq.~(\ref{M1}). The long\nobreakdash-range component stems from
the SDW component perpendicular to the $\mathbf{M}_{\text{F}}$~vector and
turns to zero at ${\alpha =0}$.

The components of~$\check{a}$ that determine the spin
current~$j_{\text{Sp}}^{x(y)}(x)$ can be presented in the same form as the
$\check{s}_{\parallel,\perp}$~matrix, i.e., ${\check{a} = \check{a}_{\parallel}
+ \check{a}_{\perp}}$. The ``longitudinal''---$\check{a}_{\parallel}$, and the
``transverse''---$\check{a}_{\perp}$, parts are related with~$\check{s}_{\parallel,\perp}$
through the simple formulas~(see Eq.~(\ref{M2b}))
\begin{align}
v n_{x}^{2} \partial_x \check{a}_{\parallel} &= 2 s_{\omega}
\hat{\tau}_{3} \cdot (|\omega| \hat{\sigma}_{0} + \mathrm{i} \mathcal{H} \hat{\sigma}_{1}) \cdot
\check{s}_{\parallel} \,,  \label{M11a} \\
v n_{x}^{2} \partial_x \check{a}_{\perp} &= 2 s_{\omega}
|\omega| \hat{\tau}_{3} \cdot \hat{\sigma}_{0} \cdot \check{s}_{\perp} \,. \label{M11b}
\end{align}

From Eqs.~(\ref{M8a}--\ref{M11b}) we find the matrix~$\check{a}$:
\begin{align}
\hat{\tau}_{3} \cdot \check{a}_{\parallel} &= -2 m \frac{\mathcal{T}_{\text{eff}}^{2}}{|n_{x}|}
\hat{\rho}_{1} \cdot \hat{\tau}_{2} \cdot \big[ \hat{\sigma}_{3} \cdot \bar{\mathcal{R}}(x) + \mathrm{i} \hat{\sigma}_{0}\cdot
\bar{\mathcal{I}}(x) \big] \cdot \cos (\alpha/2) \,, \label{M11c} \\
\hat{\tau}_{3} \cdot \check{a}_{\perp} &= -2 m \frac{\mathcal{T}_{\text{eff}}^{2}}{|n_{x}|}
\hat{\rho}_{2} \cdot \hat{\tau}_{1} \cdot \hat{\sigma}_{2} \cdot \bar{\mathcal{C}}(x) \cdot \sin (\alpha/2) \,,  \label{M11d}
\end{align}%
where
\begin{align}
\bar{\mathcal{R}}(x) &= \Re \big\lbrace \bar{J}(x) \big\rbrace \,, & \bar{\mathcal{I}}(x) &= \Im \big\lbrace \bar{J}(x) \big\rbrace \label{M10c} \,, \\
\intertext{and}
\bar{J}(x) &= \frac{\sinh (\theta_{+} x/L)}{\sinh (\theta_{+})} \,, & \bar{\mathcal{C}}(x) &= \frac{\cosh (\theta_{\omega} x/L)}{\cosh (\theta_{\omega})} \,. \label{M10d}
\end{align}
These functions are plotted in Fig.~\ref{fig:Re_Im_C}~b) and contribute to
the ``short''- and the long\nobreakdash-range components of the Green's function
in the F\nobreakdash-layer, more exactly---to the odd in~$n_x$ part of it.

As it follows from the expression for the spin current~(see Eq.~(3.13)
in~Ref.~\onlinecite{MoorVE11})), the obtained formulas for~$\check{a}_{\parallel,\perp}$
lead to zero spin current inside the F\nobreakdash-film. This result is quite
natural because the spin Josephson-like current in SDW/F/SDW junction should
be proportional to the penetration probabilities of both interfaces,
i.e.,~${j_{\text{Sp}}^{x(y)}(x) \sim \mathcal{T}_{\text{eff(L)}}^{2} \mathcal{T}_{\text{eff(R)}}^{2}}$.
In the symmetric case under consideration ${\mathcal{T}_{\text{eff(L)}}^{2} = \mathcal{T}_{\text{eff(R)}}^{2} = \mathcal{T}_{\text{eff}}^{2}}$.
In order to obtain a finite spin current, we have to calculate the matrix~$\check{a}$ in
the next order in the transmission coefficient~$\mathcal{T}_{1,2}^{2}$. We can
easily find the corrections~$\delta \check{a}_{\parallel,\perp}$ from the normalization
condition
\begin{equation}
\check{g} \cdot \check{g} = 1 \,.  \label{M12}
\end{equation}
As it follows from this equation
\begin{equation}
\check{a}_{t} \cdot (\check{g}_{0\text{F}} + \check{s}_{t}) + (\check{g}_{0\text{F}} + \check{s}_{t}) \cdot \check{a}_{t} = 0 \,,  \label{M13}
\end{equation}
where~$\check{a}_{t} = \check{a} + \delta \check{a}$ and~$\check{s}_{t} = \check{s} + \delta \check{s}$.

Therefore, for~$\delta \check{a}$ we obtain the equation
\begin{equation}
2 \check{g}_{0\text{F}} \cdot \delta \check{a} + \check{a} \cdot \check{s} + \check{s} \cdot \check{a} = 0 \,,  \label{M14a}
\end{equation}%
where ${\check{g}_{0\text{F}} = \mathrm{sgn}(\omega) \hat{\rho}_{0} \cdot \hat{\tau}_{3} \cdot \hat{\sigma}_{0}}$.
We are interested in the component~${\delta {\check{a}}}_{3}$ of the matrix~$\delta \check{a}$, which is
proportional to the matrix~$\hat{\tau}_{3}$, because only this component contributes to the spin current. This component
commutes with the matrix~$\check{g}_{0\text{F}}$.

Substituting expressions (\ref{M9a}--\ref{M10d}) into Eq.~(\ref{M14a}), we
obtain for~$\delta {\check{a}}_{3}$
\begin{equation}
\delta \check{a}_{3} = 2 \mathrm{i} s_{\omega} \frac{\big( m \mathcal{T}_{\text{eff}}^{2} \big)^{2}}{|n_{x}|}
\Re \Bigg\lbrace \frac{ \cosh \big[ (\theta_{+} - \theta_{\omega}) x/L \big]}{\sinh(\theta_{+} ) \cosh(\theta_{\omega})} \Bigg\rbrace
\sin (\alpha) \hat{\rho}_{3} \cdot \hat{\tau}_{3} \cdot \hat{\sigma}_{1} \,. \label{M15}
\end{equation}
One can write Eq.~(\ref{M15}) in the form
\begin{align}
\delta \check{a}_{3} = 2 \mathrm{i} s_{\omega} &\frac{\big(m \mathcal{T}_{\text{eff}}^{2} \big)^{2}}{|n_{x}|}
\tanh(\theta_{\omega}) \frac{\cos(\theta_{\mathcal{H}})}{\cosh^{2}(\theta_{\omega}) - \cos^{2}(\theta_{\mathcal{H}})} \label{M15a} \\
&\times \cos(\theta_{\mathcal{H}} x/L) \sin(\alpha) \hat{\rho}_{3} \cdot \hat{\tau}_{3} \cdot \hat{\sigma}_{1} \,, \notag
\end{align}
where ${\theta_{\mathcal{H}}=\mathcal{H}L / v|n_{x}|}$.

Now, the spin current can be easily calculated.

\subsection{Spin Current}

We see that the SDW/F/SDW junctions are very similar to the S/F/S Josephson
junctions. In analogy to the ``short''- and long\nobreakdash-range
components of the superconducting condensate penetrating the ferromagnet,
there are similar ``short''- and long\nobreakdash-range components of the
SDW penetrating the ferromagnet.

Nevertheless, there is an essential difference between the two systems. The
Josephson charge current~$j_{\text{J}}$ in an S/F/S junction is constant in
space in the ferromagnet. It is determined by the long-range component,
provided the magnetization~$\mathbf{M}(x)$ in the F\nobreakdash-layer is not
uniform and the thickness~$d_{\text{F}}$ of this layer is not too
small~(${d_{\text{F}} \gg \xi_{\text{F}}}$).

\begin{figure}[t]
\begin{center}
\includegraphics[width=\columnwidth]{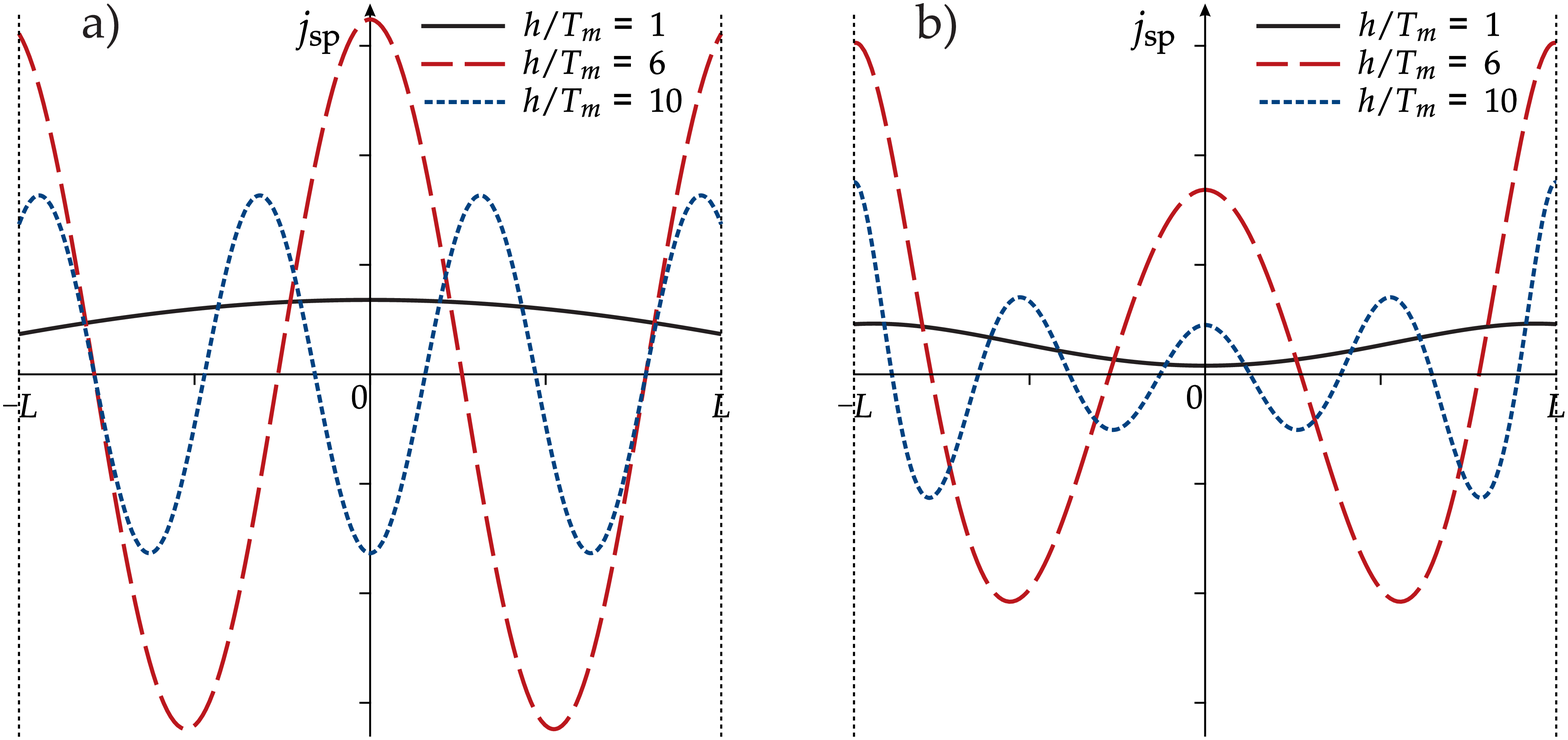}
\end{center}
\caption{{\protect\small {(Color online.) The coordinate dependence of the
spin current in the F-layer for different values of the strength of the
magnetic field~${h\equiv\mathcal{H}}$; $T_m$~is the SDW transition temperature.\\
a) ${\big(\mathcal{T}_{\text{eff}}\big)^2 \propto \delta(1-|n_x|)}$, \\
b) ${\big(\mathcal{T}_{\text{eff}}\big)^2 = \gamma_B = \text{const}}$.}}}
\label{fig:JSpx_on_x}
\end{figure}

In contrast, the spin current in an SDW/F/SDW junction is not constant in
the F\nobreakdash-layer but oscillates with a period of
the order of~$\xi _{\text{F}}$ (see~Fig.~\ref{fig:JSpx_on_x}). It is
constant in space only in the absence of the exchange field, i.e., in
SDW/N/SDW junction.

The spin current through the SDW/F/SDW junction can easily be calculated
using the formula for~$j_{\text{Sp}}$~(see Eq.~(3.13) of~Ref.~\onlinecite{MoorVE11})
\begin{equation}
j_{\text{Sp}}^{x(y)}(x) = -\mathrm{i} \frac{(2\pi T) v \nu}{8}
\mu_{B} \sum_{\omega} \mathrm{Tr} \langle (\hat{\rho}_{3} \cdot \hat{\tau}_{3} \cdot
\hat{\sigma}_{1,2} \cdot \check{a}) n_{x}^{2} \rangle \,.
\label{SpinCurrGeneral}
\end{equation}
The notation~$j_{\text{Sp}}^{x(y)}(x)$ means the spin current density with
the spin projection $x$~(resp.~$y$) flowing perpendicular to the S/F interface and
the angle brackets denote the angle averaging.

Substituting Eq.~(\ref{M15}) into Eq.~(\ref{SpinCurrGeneral}) we obtain
\begin{equation}
j_{\text{Sp}}^{x}(x) = j_{\text{c,Sp}}^{x}(x) \cdot \sin(\alpha ) \,,
\label{eqn:Spin_Current}
\end{equation}%
where
\begin{align}
j_{\text{c,Sp}}^{x}(x) &= \mu_{\text{B}} v \nu (2 \pi T)   \label{j_c(x)} \\
&\times \sum_{\omega} \Bigg\langle \frac{|{n}_{x}| m^{2} \mathcal{T}_{\text{eff}}^{4} \tanh(\theta_{\omega})
\cos(\theta_{\mathcal{H}}) \cos(\theta_{\mathcal{H}} x/L)}{\cosh^{2}(\theta_{\omega}) -
\cos^{2}(\theta_{\mathcal{H}})}  \Bigg\rangle \,. \notag
\end{align}%
One can see that the spin current oscillates in space in the F\nobreakdash-layer. It is
also seen that in the case of a normal~(nonmagnetic) metal, i.e., at ${\mathcal{H}=0}$,
the spin current is constant in space.

As follows from Eq.~(\ref{j_c(x)}), the spin current at the interfaces it is equal to
\begin{align}
j_{\text{c,Sp}}^{x}(\pm L) &= \mu_{\text{B}} v \nu (2 \pi T) \label{j_c} \\
&\times \sum_{\omega} \Bigg\langle \frac{|n|_{x} m^{2} \mathcal{T}_{\text{eff}}^{4}
\tanh(\theta_{\omega}) \cos^{2}(\theta_{\mathcal{H}})}{\cosh^{2}(\theta_{\omega}) - \cos^{2}(\theta_{\mathcal{H}})} \Bigg\rangle \,.  \notag
\end{align}

Using Eq.~(\ref{j_c(x)}) we plot the coordinate dependence of the critical spin
current with the projection of the spin onto the $x$\nobreakdash-axis in
Fig.~\ref{fig:JSpx_on_x} for different values of~$\mathcal{H}$. The critical spin
current oscillates with a period of the order of~$\xi_{\text{F,b}}$. On the left
panel we choose~${\big(\mathcal{T}_{\text{eff}}\big)^2 \propto \delta(1-|n_x|)}$
so that there is no dependence of the oscillation amplitude on~$x$; on the right panel we
choose~${\big(\mathcal{T}_{\text{eff}}\big)^2 = \gamma_B = \text{const.}}$,
so that the amplitude depends on~$x$ in a power\nobreakdash-law fashion, which
is a consequence of the integration over the angles.

Thus, we see that the SDW order parameter penetrating into the ferromagnet
contains two components---a long\nobreakdash-range component with $\mathbf{M}$~vector
perpendicular to the magnetization in the ferromagnet~$\mathbf{M}_{\text{F}}$,
and a ``short''\nobreakdash-range component with $\mathbf{M}$~vector parallel to
the $z$\nobreakdash-axis. A similar effect arises in an S/F/S junction with
the nonhomogeneous magnetization, where the long\nobreakdash-range component is the odd
triplet~(${S\neq0}$) wave function of the Cooper pairs, and the short\nobreakdash-range
component corresponds to the singlet and ${S=0}$~triplet wave function.\cite{BVErmp,PhysToday}
However, the spin current~$j_{\text{Sp}}$ is not analogous to the charge
Josephson current~$j_{\text{J}}$---it is not constant in space, but oscillates with
a period~${\sim \xi_{\text{F,b}}}$.

\section{SDW/I/SDW ``Josephson'' junction}

\subsection{Ferrell--Prange Equation}

Now, we consider a junction formed by two materials with the~SDW. These can
be iron-based pnictides separated by an insulating, normal metal or
ferromagnetic layer. In all these cases the current through the junction is
given by
\begin{equation}
j_{\text{Sp}}^{x} = j_{\text{c,Sp}}^{x} \cdot \sin(\alpha) \,,  \label{AF/AF}
\end{equation}%
where the critical spin current depends on the type of the barrier. For
example, for SDW/N/SDW junction it is equal to
\begin{equation}
j_{\text{c,Sp}}^{x} = 2 \mu_{\text{B}} v \nu (2 \pi T) W_{M0}^{2} \sum_{\omega}
\Bigg\langle \frac{|n_{x}| \mathcal{T}_{\text{eff}}^{4}}{\mathcal{E}_{M}^{2} \sinh(2 \theta_{\omega})} \Bigg\rangle \,.  \label{CrCurN}
\end{equation}
Here, the superscript~$x$ denotes the component of the spin current vector
having the spin projection on the $x$\nobreakdash-axis only. For the case of
SDW/I/SDW junction we easily obtain
\begin{equation}
j_{\text{c,Sp}}^{x} = 2 \mu_{\text{B}} v \nu (2\pi T) W_{M0}^{2}
\sum_{\omega} \Bigg\langle \frac{|n|_{x} \mathcal{T}_{\text{eff}}^{2}}{\mathcal{E}_{M}^{2}} \Bigg\rangle \,.  \label{CrCurI}
\end{equation}
Introducing ${\tilde{\Gamma} = \big\langle |n|_{x} \mathcal{T}_{\text{eff}}^{2} \big\rangle }$
we can write this equation in the case of ideal nesting as
\begin{align}
j_{\text{c,Sp}}^{x} &= \tilde{\Gamma} \nu \mu_{\text{B}} W_{M0}^{2} (2\pi T)
\sum_{\omega} \mathcal{E}_{M}^{-2}  \label{Amb_Baratoff} \\
&= \frac{\mu_{B} W_{M0}}{e^2 \tilde{R}} \tanh\frac{W_{M0}}{2 \pi T} \,, \notag
\end{align}
which corresponds to the Ambegaokar--Baratoff formula for the critical Josephson
current with $\Delta $ replaced by~$W_{M0}$ and the ``interface resistance per
unit area''~$\tilde{R}$ is related to the effective transmission
coefficient---${\tilde{R} \propto 1/\tilde{\Gamma}}$.

Equation~(\ref{AF/AF}) for the spin ``Josephson'' current in an
AF/I/AF junction has been obtained in Ref.~\onlinecite{Tremblay} by another method.

Now, we derive an equation similar to the Ferrell--Prange\cite{Ferrell}
equation. We consider a junction with SDW leads in the form of thin films.
The thickness~$d$ of the films must be less than a characteristic length,
over which the SDW vector~$\mathbf{M}$ restores its preferable direction.
This direction is determined by anisotropy effects.

\begin{figure}[t]
\begin{center}
\includegraphics[width=0.4\textwidth]{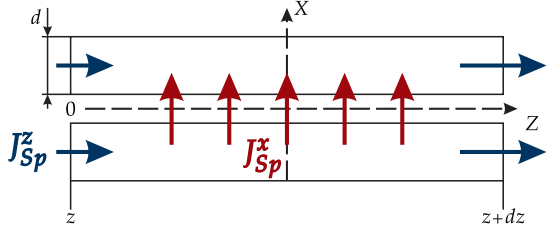}
\end{center}
\caption{{\protect\small {(Color online.) A scheme for derivation of the
Ferrell--Prange equation.}}}
\label{fig:Ferrell_Prange}
\end{figure}
\begin{figure}[t]
\begin{center}
\includegraphics[width=0.4\textwidth]{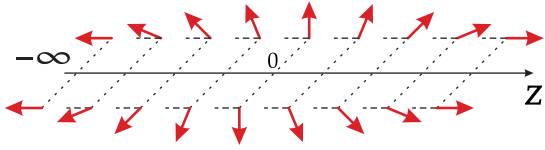}
\end{center}
\caption{{\protect\small {(Color online.) Dependence of the magnetization
direction in the ``fluxon'' with~$z$. The mutual orientation in the leads
changes from~$0$ to~$2\protect\pi $.}}}
\label{fig:Spin_Fluxon}
\end{figure}

The idea of the derivation is the same as in Ref.~\onlinecite{Ferrell}.
Considering the region near the boundary~(cf.~Fig.~\ref{fig:Ferrell_Prange})
we use the continuity equation for the spin~$\mathbf{M}^{(x)}$,
\begin{equation}
\partial_{t} \mathbf{M}^{(x)} = - \mathrm{div} \mathbf{j}_{\text{Sp}} \,,
\label{ContinM_x}
\end{equation}%
which can be derived easily from Eq.~(3.5) of Ref.~\onlinecite{MoorVE11}. Using
the Green's functions~$\check{g}$ unperturbed by the proximity effect (see
Eq.~(\ref{M1})), one can show that ${\mathbf{M}^{(x)}=0}$.
From the continuity equation we get
\begin{equation}
d \cdot \lbrack \partial_{z} j_{\text{Sp}}^{z}(z)_{|_{x=+0}} - \partial_{z} j_{\text{Sp}}^{z}(z)_{|_{x=-0}} \rbrack = 2 \cdot j_{\text{Sp}}^{x} \,.  \label{FPcont}
\end{equation}

The formula for the spin current $j_{\text{Sp}}^{x}$ in an SDW lead with
the magnetization orientation varying in space can be easily obtained from
Eq.~(\ref{M2}) in the same way as the Meissner current is found in a clean
superconductor~(see, e.g., Ref.~\onlinecite{Kopnin}). It has the form~(see
Appendix~\ref{Ferrell_Prange_Appendix})
\begin{equation}
j_{\text{Sp}}^{z}(z)_{|_{x=\pm 0}} = - \frac{1}{6} \frac{\partial \alpha}{\partial z}
v^{2} \nu \mu_{B} (2\pi T) W_{M0}^{2} \sum_{\omega} {\mathcal{E}_{M}^{-3}} \,.  \label{SpinCurrUn}
\end{equation}
This equation formally coincides with the corresponding formula for the
supercurrent in a clean superconductor if one replaces ${W_{M0} \rightarrow \Delta}$,
${\alpha \rightarrow \chi}$ and ${\mu_{B} \rightarrow 2e}$, where $\chi$ is the
phase in the superconductor.

Substituting Eq.~(\ref{AF/AF}) into Eq.~(\ref{FPcont}) we arrive at the final
result
\begin{equation}
d l_{\text{c}} \partial_{z}^{2} \alpha = \sin(\alpha) \,,  \label{FPfinal}
\end{equation}%
where~$l_{\text{c}}$ is a characteristic length inversely proportional to the
critical current~$j_{c,\text{Sp}}$. It equals
\begin{align}
l_{\text{c}} &= \left( \sum_{\omega} \Bigg\langle \frac{|n_{x}| \mathcal{T}_{\text{eff}}^{4}}{\mathcal{E}_{M}^{2} \sinh(2 \theta_{\omega})} \Bigg\rangle \right)^{-1} \frac{\hbar v}{3} \sum_{\omega}{\mathcal{E}_{M}^{-3}} \\
\intertext{in the case of SDW/N/SDW junctions and}
l_{\text{c}} &= \left( \sum_{\omega} \Bigg\langle \frac{|n|_{x} \mathcal{T}_{\text{eff}}^{2}}{\mathcal{E}_{M}^{2}} \Bigg\rangle \right)^{-1} \frac{\hbar v}{3} \sum_{\omega}{\mathcal{E}_{M}^{-3}}
\end{align}
for SDW/I/SDW junctions.

Equation~(\ref{FPfinal}) is the analog of the Ferrell--Prange
equation for a junction composed of two materials with the SDW and separated
by some barrier (a thin insulating or normal metal layer).

One can easily find a solution similar to the one describing a fluxon in the
Josephson tunnel junctions. It has the form
\begin{equation}
\alpha = 4 \arctan \left[ \exp (z/l_{0})\right] \,,  \label{Fluxon}
\end{equation}%
where the size of the ``fluxon'' is given by ${l_{0} = \sqrt{d l_{\text{c}}}}$. The
mutual orientation changes along the $z$\nobreakdash-axis from~$0$ to~$2\pi$
and has the value~$\pi$ in the core of the ``fluxon'', cf.~Fig.~\ref{fig:Spin_Fluxon}.
In the case of the Josephson fluxon one has a circulating
charge current---in our case, the spin current circulates in the ``fluxon''.

\subsection{Finite Detuning}

Up to now, we assumed the ideal nesting, ${\delta \mu =0}$. In the case
${\delta \mu \neq 0}$ we find the quasiclassical Green's function in the SDW
system by inverting the matrix~$\check{G}$ in Eq.~(3.3) of Ref.~\onlinecite{MoorVE11}.
It turns out that~$\check{g}$ is a linear combination of six ``basis matrices''
\begin{equation}
\check{g}_{\text{L/R}}=\check{g}_{030}\pm \check{g}_{122}+\check{g}_{123}\pm
\check{g}_{212}+\check{g}_{213}+\check{g}_{300}\,,  \label{eqn:QGF_in_AF}
\end{equation}%
where the subscripts L and R denote the left and right SDW systems,
respectively, corresponding to the setup depicted in Fig.~\ref{fig:Setup}
with the F\nobreakdash-layer replaced by a very thin insulating barrier. The
six functions~$\check{g}_{ijk}$ are
\begin{align}
\check{g}_{030}& =|\chi _{+}|^{-2}\left( \omega \cdot \Re \left\{ \chi
_{+}\right\} +\delta \mu \cdot \Im \left\{ \chi _{+}\right\} \right) \cdot
\hat{\rho}_{0}\cdot \hat{\tau}_{3}\cdot \hat{\sigma}_{0}\,, \\
\check{g}_{122}& =\mathrm{i}|\chi _{+}|^{-2}W_{M0}\cdot \Im \left\{ \chi
_{+}\right\} \sin (\alpha/2)\cdot \hat{\rho}_{1}\cdot \hat{\tau}_{2}\cdot
\hat{\sigma}_{2}\,, \\
\check{g}_{123}& =|\chi _{+}|^{-2}W_{M0}\cdot \Re \left\{ \chi _{+}\right\}
\cos (\alpha/2)\cdot \hat{\rho}_{1}\cdot \hat{\tau}_{2}\cdot \hat{\sigma}%
_{3}\,, \\
\check{g}_{212}& =|\chi _{+}|^{-2}W_{M0}\cdot \Re \left\{ \chi _{+}\right\}
\sin (\alpha/2)\cdot \hat{\rho}_{2}\cdot \hat{\tau}_{1}\cdot \hat{\sigma}%
_{2}\,, \\
\check{g}_{213}& =\mathrm{i}|\chi _{+}|^{-2}W_{M0}\cdot \Im \left\{ \chi
_{+}\right\} \cos (\alpha/2)\cdot \hat{\rho}_{2}\cdot \hat{\tau}_{1}\cdot
\hat{\sigma}_{3}\,, \\
\check{g}_{300}& =\mathrm{i}|\chi _{+}|^{-2}\left( \omega \cdot \Im \left\{
\chi _{+}\right\} -\delta \mu \cdot \Re \left\{ \chi _{+}\right\} \right)
\cdot \hat{\rho}_{3}\cdot \hat{\tau}_{0}\cdot \hat{\sigma}_{0}\,,
\end{align}%
where ${\chi_{+} = \sqrt{W_{M0}^{2} + (\omega + \mathrm{i} \delta \mu )^{2}}}$.

The detuning parameter~$\delta \mu $ can be written in the form
${\delta \mu = \mu_{0} + \mu_{\phi} \cos(2\phi)}$ with $\mu_{0}$ describing
the relative size mismatch of the ``hole'' and ``electron'' pockets and
$\mu_{\phi}$ describes the ellipticity of the ``electron'' pocket, while
the ``hole'' pocket is assumed to be a circle.\cite{Chubukov09,Chubukov10,Schmalian10}

From the quasiclassical Green's function, Eq.~(\ref{eqn:QGF_in_AF}), we can
calculate the dependence of various quantities of interest on~$\delta\mu$.

We begin with the density of states~(DoS). The DoS is given by the formula
\begin{equation}
\nu(\epsilon) = \frac{1}{2} \Re \bigl\langle \mathrm{Tr} \bigl( \hat{\rho}_0
\cdot \hat{\tau}_{3} \cdot \hat{\sigma}_{0} \cdot \check{g}(\epsilon) \bigr) %
\bigr\rangle \,,  \label{eqn:DoS}
\end{equation}
where the angle brackets mean the averaging over the momentum directions
\begin{equation}
\bigl\langle\bigl(\dots\bigr)\bigr\rangle = \int \frac{\mathrm{d} \Omega}{4
\pi} \, \bigl(\dots\bigr) \,.
\end{equation}

Inserting the expression for the quasiclassical Green's function into Eq.~(\ref{eqn:DoS})
we obtain
\begin{equation}
\nu (\epsilon) = \frac{1}{2} \Re \left\langle \frac{\epsilon + \delta \mu }{%
\sqrt{(\epsilon +\delta \mu )^{2}-W_{M0}^{2}}}+\frac{\epsilon -\delta \mu }{%
\sqrt{(\epsilon -\delta \mu )^{2}+W_{M0}^{2}}}\right\rangle \,.
\end{equation}%
This result was obtained earlier in Refs.~\onlinecite{Chubukov10,Schmalian10}.

\begin{figure}[t]
\begin{center}
\includegraphics[width=0.3\textwidth]{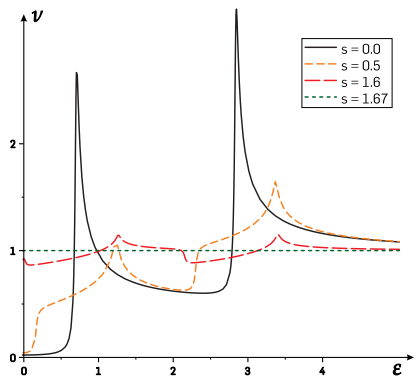}
\end{center}
\caption{{\protect\small {(Color online.) DoS for different values of ${s=%
\protect\mu_{\protect\phi}/\protect\mu_{0}}$ at ${\protect\mu_{0}/T_{\text{s}%
} \simeq 1.07}$.}}}
\label{fig:DoS_AF}
\end{figure}

\begin{figure}[t]
\begin{center}
\includegraphics[width=0.3\textwidth]{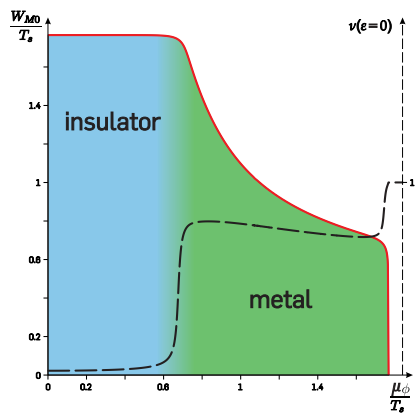}
\end{center}
\caption{{\protect\small {(Color online.) $W_{M0}(\protect\mu_{\protect\phi%
}) $ at ${\protect\mu_{0}/T_{\text{s}} \simeq 1.07}$; black dashed curve
shows the dependence of DoS on $\protect\mu_{\protect\phi}$ at zero energy.}}
}
\label{fig:Wm_on_mu_phi}
\end{figure}

We plot the DoS for different values of the ratio ${s=\mu_{\phi}/\mu _{0}}$
in Fig.~\ref{fig:DoS_AF} choosing a constant ${\mu_{0} \simeq 1.07T_{\text{s}}}$,
where $T_{\text{s}}$ is the SDW order transition temperature. Note that
there is a range of the values of~$s$ where an energy gap appears in the~DoS,
indicating an insulating phase of the system. We can trace the
transition to the metal considering the value of the DoS at zero energy,~$\nu(0)$,
as a function of the parameter~$\mu_{\phi}$. In Fig.~\ref{fig:Wm_on_mu_phi}
we plot the dependence of the order parameter~$W_{M0}$ on~$\mu_{\phi}$ and,
in the same plot, show the dependence~$\nu(0)$ on~$\mu_{\phi}$~(dashed line).
One can observe a rather steep increase of~$\nu(0)$ in the region around~$0.7T_{\text{s}}$.

The dependence of the SDW order parameter~$W_{M0}$ on~$\mu_{\phi}$ is
calculated from the self\nobreakdash-consistency equation, Eq.~(3.19) of
Ref.~\onlinecite{MoorVE11}, which in addition has to be averaged over the momentum
directions. It reduces to the form (see also~Ref.~\onlinecite{Chubukov10})
\begin{equation}
\ln \left( T/T_{\text{s}}\right) = 2 \pi T \sum_{\omega > 0} \Re \left\langle
\chi_{+}^{-1} - \omega^{-1} \right\rangle \,.  \label{eqn:Self_Consistency}
\end{equation}

It is of special interest\cite{Sonin} to calculate the spin current through
the SDW/I/SDW system. Using the expression Eq.~(\ref{eqn:QGF_in_AF}) for the
quasiclassical Green's functions in the left and right leads, and inserting
it into the expression for the spin current, Eq.~(3.13) of~Ref.~\onlinecite{MoorVE11},
we obtain the familiar form of the Josephson-like current
\begin{equation}
j_{\text{Sp}} = j_{\text{c,Sp}} \cdot \sin(\alpha)
\label{AF/AF_finite_mu}
\end{equation}%
for the spin current through the SDW/SDW interface which has only the spin
projection on the $x$\nobreakdash-axis. The critical coefficient~$j_{\text{c,Sp}}$
is dependent on~$\delta \mu$
\begin{equation}
j_{\text{c,Sp}} = 2 \mu_{\text{B}} v \nu W_{M0}^{2} (2\pi T)
\Bigg\langle |n_x| \mathcal{T}_{\text{eff}}^2 \sum_{\omega > 0} \frac{\Re \left\{ \chi_{+} \right\}^{2}
- \Im \left\{ \chi_{+} \right\}^{2}} {|\chi_{+}|^{4}} \Bigg\rangle \,.
\label{eqn:Spin_Current_on_Mu}
\end{equation}

\begin{figure}[t]
\begin{center}
\includegraphics[width=0.4\textwidth]{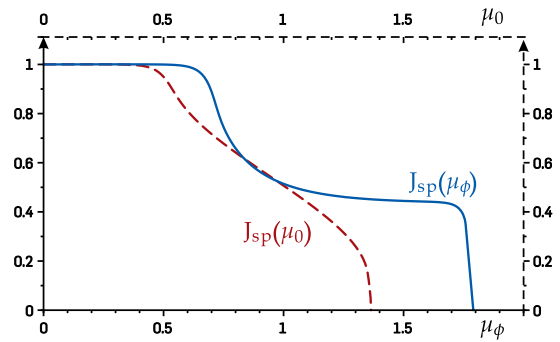}
\end{center}
\caption{{\protect\small {(Color online.) Spin current on $\protect\mu_{0}$
resp.\ $\protect\mu_{\protect\phi}$.}}}
\label{fig:Spin_Current_on_mu}
\end{figure}

In Fig.~\ref{fig:Spin_Current_on_mu} we show the dependence of the critical
spin current on the two parameters~$\mu_0$ and~$\mu_{\phi}$, choosing in
each case the other parameter constant---when plotted as a function of~$\mu_0$,
we set ${\mu_{\phi} = 1.26 T_{\text{s}}}$; plotting the spin current as a
function of~$\mu_{\phi}$ we set ${\mu_{0} = 1.07 T_{\text{s}}}$.

One can see from Fig.~\ref{fig:Spin_Current_on_mu} and Fig.~\ref%
{fig:Wm_on_mu_phi} that in the insulating phase the spin current is not
zero. In contrary, it is at its maximum. This addresses the question about
the possibility of the appearance of the spin current in an insulator (see
discussion on p.~12 of~Ref.~\onlinecite{Sonin}).

Analyzing the expression~(\ref{eqn:Spin_Current_on_Mu}) for the critical
spin current we see that there is a possibility of a sign change. At small
temperatures, the spin current changes sign if, e.g., the condition
${W_{M0} \lesssim \mu_{0} - \mu_{\phi}}$ holds with ${\mu_{\phi} < \mu_{0}}$
at the same time.

We note the similarity of the self-consistency equation~(\ref%
{eqn:Self_Consistency}) with the self-consistency equations presented in
Refs.~\onlinecite{LO,FF}. Setting ${\mu_{\phi}=0}$ we can map the problem of
finding the dependence of $W_{M0}$ on~$\mu_{0}$ onto the problem of finding
the dependence of the superconducting order parameter on a strong exchange
field resulting in the so-called Larkin--Ovchinnikov--Fulde--Ferrell~(LOFF)
state, which is characterized by the spatial dependence of~$\Delta$. Note
that the possibility of the LOFF-like state in pnictides was noted earlier
by Gor'kov and Teitel'baum.\cite{Gorkov2010}

\section{Discussion}

We have considered Josephson-like junctions of the SDW/F/SDW and SDW/I/SDW
types and calculated the spin current~$j_{\text{Sp}}$ in these systems. In
both the cases the dependence of~$j_{\text{Sp}}$ on the angle~$\alpha$
between the mutual orientations of the magnetization vectors~$\mathbf{M}$ in
the right and left leads is given by~Eq.~(\ref{AF/AF}). The critical current
density~$j_{\text{Sp,c}}$ is determined by the interface transparencies.

We emphasize the analogy between the SDW/F/SDW junctions and the Josephson
S/F/S junctions with a nonhomogeneous magnetization. In both the systems
there are ``short''- and long\nobreakdash-range components of the order parameters that
penetrate the ferromagnet. In the ballistic systems under consideration the
``short''\nobreakdash-range SDW component penetrating the F\nobreakdash-layer oscillates
with a period of the order of~${\xi_{\text{F,b}} = \hbar v/\mathcal{H}}$. It stems from the SDW
component parallel to the magnetization vector~$\mathbf{M}_{\text{F}}$ in
the ferromagnet. The SDW component normal to the $\mathbf{M}_{\text{F}}$
vector penetrates the ferromagnet on a long distance of the order
of~${\xi_{\text{N,b}} = \hbar v / 2 \pi T}$. The penetration depth of this component does not
depend on the exchange field~$\mathcal{H}$, and, in this sense, this
component is analogous to the odd triplet long\nobreakdash-range component in S/F/S
junctions with a nonhomogeneous magnetization.

However, there is an essential difference between the spin current in
SDW/F/SDW junctions~$j_{\text{Sp}}(x)$ and the Josephson charge
current~$j_{\text{J}}$ in S/F/S junctions. Whereas the Josephson current~$j_{\text{J}}$,
as it should be, does not depend on~$x$ (${\mathrm{div}j_{\text{J}} = \partial_{x} j_{\text{J}}=0}$),
the spin current~$j_{\text{Sp}}(x)$ oscillates in
space with a period of the order~${\xi_{\text{F,b}}}$. This behavior is a
consequence of the fact that the spin current arises due to the interference
between the ``short''- and long\nobreakdash-range component of the SDW in the ferromagnet.

The spin current in SDW/I/SDW junctions depends on the angle~$\alpha $ in the
same way as in the SDW/F/SDW junctions, i.e., ${j_{\text{Sp}} \sim \sin(\alpha)}$.
We have calculated the dependence of~$j_{\text{Sp}}$ in
an SDW/I/SDW junction on the parameter~$\delta \mu $ characterizing the
deviation from the ideal nesting. At ${\delta \mu = 0}$, an energy gap~$W_{M0}$
opens in the excitation spectrum in the leads with the SDW~(see
Fig.~\ref{fig:Wm_on_mu_phi}). In a certain region of the parameter~$\delta \mu$,
the energy gap disappears and the leads of the junction become metals~(or
semimetals). The spin current exists in both the cases; it turns to zero at
the Neel temperature when~${W_{M0} \rightarrow 0}$.

Furthermore, we have derived the Ferrell--Prange equation for the Josephson-like
junctions with the SDW and found a solution describing a localized spin current
distribution analogously to the fluxon in a tunnel S/I/S Josephson junction.
The characteristic length of the ``fluxon'', ${l_0 = \sqrt{dl_{\text{c}}}}$,
is determined by the barrier transmittance and the thickness~$d$ of the leads
with the~SDW. In the case of bulk leads the thickness~$d$ should be replaced
by a characteristic length over which the vector~$\mathbf{M}$ of the SDW
restores its favorable direction in the bulk, which is determined by the
anisotropy effects.

\subsection{Acknowledgements}

The authors are grateful to A.~Chubukov and I.~Eremin for useful remarks and
discussions. We are also indebted to G.~E.~Volovik for useful comments. We thank SFB~491 for financial support.

\appendix

\section{Model Boundary Condition}
\label{Boundary_Conditions}

Here, we generalize the boundary conditions for quasiclassical
Green's functions to the case of a two-band material with the SDW. It is not
in the scope of the paper to present a rigorous derivation of the boundary
conditions with account for all the possible processes accompanying the passage
of electrons through the interface. We ignore the interband transitions and also
the spin-flip processes. In addition, we restrict ourselves to the case of a low
transmittance coefficient. Thus, the derived boundary conditions, to some
extent, are phenomenological. However, in a single\nobreakdash-band case they coincide
with the boundary conditions derived by Zaitsev\cite{Zaitsev} and in case of
two\nobreakdash-band material they are compatible with the boundary conditions
used in Ref.~\onlinecite{Brinkman04}. Note also that the exact form of the boundary
conditions is not important for our qualitative conclusions such as the appearance
of the ``short''- and long\nobreakdash-range SDW components in the ferromagnet.

In the case of weak penetration and single\nobreakdash-band materials, the boundary
conditions for the Green's function~$\check{g}$ have the form of Eq.~(\ref{M3}),
in which the matrix~$\check{\mathcal{T}}$ should be replaced by a scalar
$\mathcal{T}$~(see Ref.~\onlinecite{Zaitsev}). The Green's function on
the left-hand side of this equation determines the charge current through the
interface. It equals the commutator multiplied by~$\mathcal{T}^{2}$ and
consisting of the symmetric in momentum space Green's functions. This part
of~Eq.~(\ref{M3}), which of course also determines the charge current, can
be obtained by using the tunneling Hamiltonian method~(see, e.g, Ref.~\onlinecite%
{Volkov75}). In terms of the operators~$\check{C}_{\text{l(r)}}$ and~$\check{C}_{\text{l(r)}}^{\dagger}$
introduced in Ref.~\onlinecite{MoorVE11} the tunneling Hamiltonian can be
written as follows
\begin{equation}
\mathcal{H}_{\text{T}} = \sum_{\mathbf{k},s,s^{\prime}} \left(
    \mathcal{ \check{\mathcal{T}}} \check{C}_{\text{l}}^{\dagger} \check{C}_{\text{r}} + \text{c.c.} \right) \,,  \label{Ap1}
\end{equation}%
where $\check{\mathcal{T}} = \big( \mathcal{T}_{3} \hat{\rho}_{3} + \mathcal{T}_{0} \hat{\rho}_{0} \big)
\cdot \hat{\tau}_{3} \cdot \hat{\sigma}_{0}$ and the
coefficients~$\mathcal{T}_{0,3}$ are defined in Eq.~(\ref{M3}) in terms of
the matrix elements~$T_{1,2}$ describing the electron transition through
the interface in each band. The Green's function~$\hat{G}$ satisfies the
equation (see, e.g, Ref.~\onlinecite{Volkov75})
\begin{equation}
\hat{G}_{\text{l}} = \hat{G}_{0,\text{l}} + \hat{G}_{0,\text{l}} \cdot \hat{\Sigma}_{\text{T}} \cdot \hat{G}_{\text{l}} \,,  \label{Ap2}
\end{equation}%
where ${\hat{\Sigma}_{\text{T}} = \mathcal{\check{\mathcal{T}}} \hat{G}_{\text{r}} \mathcal{\check{\mathcal{T}}}}$.
This means that the self\nobreakdash-energy part has formally
the same form as in the case of the impurity scattering, but the Green's
function in~$\hat{\Sigma}_{\text{T}}$ describes another~(right) lead in the
equation for the Green's function~$\hat{G}_{\text{l}}$ of the left lead. The
charge current through the interface is determined by the trace of a commutator
\begin{align}
\mathrm{Tr} \big\{ \hat{\rho}_3 \cdot \hat{\tau}_3 \cdot \hat{\sigma}_0 \cdot \big[ \hat{\Sigma}_{\text{T}} \,, \hat{G}_{\text{l}} \big] \big\} =
\mathrm{Tr} \big\{ \hat{\rho}_3 \cdot \hat{\tau}_3 \cdot \hat{\sigma}_0 \cdot \big[ \check{\mathcal{T}} \hat{G}_{\text{r}} \mathcal{\check{\mathcal{T}}} \,, \hat{G}_{\text{l}} \big] \big\} \,.  \label{Ap3}
\end{align}
The expression in the curly brackets expressed in terms of the
quasiclassical Green's functions appears in the right-hand side of~Eq.~(\ref{M3}).

\section{Spin Current in a Bulk Material with the~SDW}
\label{Ferrell_Prange_Appendix}

In order to obtain the expression for the spin current in the left~(right)
lead flowing parallel to the interface, we assume that the angle~$\alpha$
varies in space slowly so that the change of~$\alpha$ on a short coherence
length~${\xi_{\text{SDW}} = \hbar v/W_{M0}}$ is small. The characteristic
length~$\xi_{\alpha}$ of the $\alpha$~variation is inversely proportional to the
critical current~$j_{\text{c,Sp}}^{x}$ (see below) and is much larger than
$\xi_{\text{SDW}}$. If the angle~$\alpha$ is spatially constant, then the Green's
function~${\check{g} \equiv \check{g}_{\alpha}}$ obeys the equation
\begin{equation}
\left[ \check{\Lambda} \,, \check{g}_{\alpha} \right] = 0 \,. \label{A1}
\end{equation}
The solution of this equation is given by Eq.~(\ref{M1}) and may be
represented in the form
\begin{equation}
\check{g}_{\alpha} = \check{\Lambda} / \mathcal{E}_{M} = \check{S} \check{g}_{0} \check{S}^{\dagger} \,,  \label{A2}
\end{equation}%
where the matrix ${\check{g}_{0} = \big\{ \omega_{n} \hat{\tau}_{3} + (W_{M0} / \mathcal{E}_{M})
\hat{\rho}_{1} \cdot \hat{\tau}_{2} \cdot \hat{\sigma}_{3} \big\}}$ does not
depend on~$\alpha$. The unitary matrix~$\check{S}$ is defined as
${\check{S} = \cos(\alpha /2) + \mathrm{i} \sin(\alpha /2) \hat{\rho}_{3}
\cdot \hat{\tau}_{3} \cdot \hat{\sigma}_{1}}$. If the angle~$\alpha(z)$ slowly
depends on the coordinate along the interface, we have to add a
correction~$\delta \check{g}_{\alpha}$ to the matrix~$\check{g}_{\alpha}$ which
depends on~$n_{x}$---${\delta \check{g}_{\alpha } = n_{x} \check{a}_{\alpha}}$.
The matrix~$\check{a}$ satisfies the equation
\begin{equation}
v \big( \check{S}^{\dagger} \cdot \partial_z \check{S} \cdot \check{g}_{0} + \check{g}_{0} \cdot \partial_z \check{S}^{\dagger} \cdot \check{S} \big) + 2 \mathcal{E}_{M} \check{g}_{0} \cdot \check{a}_{0} = 0 \,,  \label{A3}
\end{equation}%
where the matrices~$\check{a}_{\alpha}$ and~$\check{a}_{0}$ are
related in the same way as matrices~$\check{g}_{0}$ and~$\check{g}_{\alpha}$, i.e.,
${\check{a}_{\alpha} = \check{S} \check{a}_{0} \check{S}^{\dagger}}$.
Here, we used
\begin{equation}
\check{g}_{0} \cdot \check{a}_{0} + \check{a}_{0} \cdot \check{g}_{0} = 0 \,, \label{A4}
\end{equation}
which follows from the normalization condition~Eq.~(\ref{M12}). One easily finds
\begin{equation}
\check{S}^{\dagger} \partial_z \check{S} = (\mathrm{i}/2) \partial_z \alpha \hat{\rho}_{3} \cdot \hat{\tau}_{3} \cdot \hat{\sigma}_{1} \,. \label{A5}
\end{equation}

Combining Eqs.~(\ref{A3}),~(\ref{A5}) and the normalization
condition~${\check{g}_{0} \cdot \check{g}_{0} = 1}$, we find
\begin{equation}
\check{a}_{0} = -v (\mathrm{i}/2) \partial_z \alpha W_{M0} \mathcal{E}_{M}^{-3}
\big( -\mathrm{i} \omega \hat{\rho}_{0} \cdot \hat{\tau}_{3} \cdot \hat{\sigma}_{2} +
W_{M0} \hat{\rho}_{3} \cdot \hat{\tau}_{3} \cdot \hat{\sigma}_{1} \big) \,,  \label{A6}
\end{equation}
and using Eq.~(\ref{SpinCurrGeneral}), we obtain the~Eq.~(\ref{SpinCurrUn}) for~$j_{\text{Sp}}^{z}(z)$.

\end{document}